\documentclass[pre,showpacs,preprint]{revtex4}
\usepackage{amssymb,graphicx,amsbsy}

\begin{document}

\title{Non-Kosterlitz-Thouless transitions for the $q$-state clock models}
\author{Seung Ki Baek}
\email[Corresponding author, E-mail: ]{garuda@tp.umu.se}
\author{Petter Minnhagen}
\affiliation{Department of Physics, Ume{\aa} University, 901 87 Ume{\aa}, Sweden}

\begin{abstract}
The $q$-state clock model with the cosine potential has a single phase
transition for $q\leq4$ and two transitions for $q\geq5$. It is shown by
Monte
Carlo simulations that the helicity modulus for the five-state clock model
($q=5$) does not vanish at the high-temperature transition. This is in
contrast to the clock models with $q\geq6$ for which the helicity modulus
vanishes. This means that the transition for the five-state clock model
differs from the Kosterlitz-Thouless (KT) transition. It is also shown that
this change in the transition is caused by an interplay between the number
of angular directions and the interaction potential: by slightly modifying
the interaction potential, the KT transition for $q=6$ turns into the same
non-KT transition. Likewise, the KT transition is recovered for $q=5$ when
the Villain potential is used. Comparisons with other clock-model results
are made and discussed.
\end{abstract}

%05.70.Fh Phase transitions: general studies
%64.60.Cn Order-disorder transformations; statistical mechanics of model systems
%75.40.Cx Static properties (order parameter, static susceptibility, heat capacities, critical exponents, etc.) 

\maketitle

\section{Introduction}

The research on two-dimensional (2D) continuous phase transition has a long
history. One of the corner stones was Onsager's solution of the 2D Ising
model~\cite{domb} which led to the understanding of the connection between a
broken symmetry and the universality of a continuous phase
transition~\cite{domb}. Another corner stone was the discovery of a 2D
topological phase transition by Kosterlitz and Thouless (KT)~\cite{kt,kos}. This
topological transition does not involve any broken symmetry and has hence a
completely different character. The $q$-state clock models have the
interesting feature that both of these types of phase transitions are possible
and to some extent interfere with each other. Partly because of this, the
phase diagram for the $q$-state clock model is a longstanding question with
as yet no clear consensus, both as regarding the number of phase transitions
and the character of the phase
transitions~\cite{jkkn,elitzur,domany1,cardy,rujan,roomany,lapilli,hwang}. The
overall feature, for which there is full consensus, is that for $q\leq4$,
there is only one transition, whereas for $q\geq q_c$ there are two
transitions. However, the value of $q_c$ has been suggested to be both
$q_c=5$ as in Refs.~\cite{jkkn,elitzur,domany1,cardy} and $q_c=6$ as in
Refs.~\cite{rujan,roomany}.
The most common view is that $q_c=5$ and that the two phase transitions for
$q\geq 5$ are of the KT type~\cite{jkkn,elitzur,domany1,cardy}.
If there is only one transition for $q=5$ as suggested in
Refs.~\cite{rujan,roomany}, it should be
discontinuous~\cite{domany1}. More recently, it has been suggested that the
upper transition for the six-state clock model is not the KT
transition~\cite{lapilli,hwang}. Quite to the contrary, however, this
transition was later found to be of the KT type in Ref.~\cite{baek2}.

In the present paper, we reinvestigate the transitions for $q=4,5$, and $6$,
using Monte Carlo (MC) simulations. It is found that the five-state clock
model has two separate transitions. However, the upper transition is not a
pure KT transition, since the helicity modulus does not vanish at the
transition. It is also found that if the interaction potential is changed to
the Villain potential, then the upper transition does become a pure KT
transition. In a similar way, it is found that the six-state clock model
has the KT transition and that
it turns into a non-KT transition when the potential
is slightly changed. If the potential is further changed the upper and lower
phase transitions coalesce into one and become discontinuous. Thus both the
number of phase transitions and the character of the transitions for a
clock model with $q\geq5$ depend on the precise shape of the potential.
This work is organized as follows: Sec.~\ref{sec:clock} introduces the spin
models and statistical quantities we are investigating. These
models have somewhat different interaction potentials between spins but have
the same symmetry and quadratic leading term. We present our
main numerical results in Sec.~\ref{sec:mc} and summarize them in
Sec.~\ref{sec:con}.

\section{Clock Models and Helicity Modulus}
\label{sec:clock}

The $q$-state clock model is, like the Ising model, a model of interacting
spins where the interaction energy between two neighboring spins is
determined by the difference in spin angles. Hence we begin with a
Hamiltonian derived from a spin-interaction potential $U$ as
\begin{equation}
H = \sum_{\left< ij \right>} U(\theta_i - \theta_j),
\label{eq:h}
\end{equation}
where $\theta_i$ is the $i$th spin angle, and the sum runs over all the
nearest-neighbor pairs. We will here consider the case when the lattice is
a 2D $L\times L$ square lattice. The interaction potential for the clock model
is given by
\begin{equation}
U(\phi) = V(\phi) \equiv -J \cos \phi,
\label{eq:clock}
\end{equation}
where $J$ is the coupling constant. Each spin can only have $q$ discrete
directions, $\theta = 0, \frac{2\pi}{q}, \ldots, \frac{2\pi (q-1)}{q}$. This
means that the $q$-state clock model has a discrete $Z_q$ symmetry. In case
of the Ising model, $q=2$, the possible spin directions are $\theta = 0$ and
$\pi$, corresponding to the usual spin-up and spin-down directions, and the
symmetry is $Z_2$. In the limit $q\rightarrow\infty$, the spin angles
$\theta_i$ becomes continuous and the Hamiltonian recovers a full
$U(1)$ symmetry. This limiting model is called the 2D $XY$ model and is the
prototype of a system exhibiting the KT transition.

\begin{figure}
\includegraphics[width=0.45\textwidth]{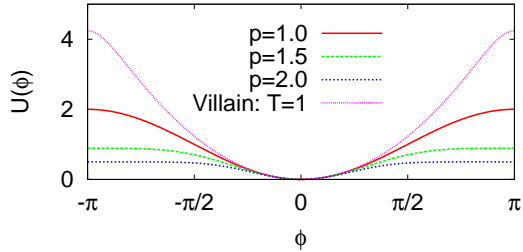}
\caption{(Color online) Interaction potentials in Eq.~(\ref{eq:p}) at
various $p$ values, compared with the Villain approximation,
Eq.~(\ref{eq:villain}).}
\label{fig:pot}
\end{figure}

The critical properties of a continuous phase transition is to a large extent
determined by the symmetry and the dimension of the model. Thus it is
reasonable to assume that a small change in the interaction potential
$V(\phi)$ will not matter as long as the $Z_q$ symmetry is preserved. Such a
change is given by the Villain approximation~\cite{villain},
\begin{equation}
U(\phi) = V_{\rm Villain}(\phi) \equiv -\frac{1}{\beta} \ln
\left\{ \sum^{\infty}_{n=-\infty}
\exp \left[ -\beta(\phi-2\pi n)^2/2 \right) \right\},
\label{eq:villain}
\end{equation}
where $\beta$ is the inverse of temperature $T$ in units such that the
Boltzmann constant is unity.
Figure~\ref{fig:pot} compares the clock-model potential $V(\phi)$ with the
Villain approximation. The point with the Villain approximation is that this
approximation makes theoretical analysis more tractable~\cite{jkkn} and it
has often been tacitly assumed that the phase diagram and phase-transition
properties obtained within the Villain approximation should be universal and
also valid for the clock-model potential,
Eq.~(\ref{eq:clock})~\cite{jkkn,elitzur}.

In order to investigate the sensitivity to the precise
functional form $V(\phi)$ in a systematic way, one can use the following
parameterization~\cite{domany,baek1}
\begin{equation}
U(\phi) = V_p(\phi) \equiv \frac{2J}{p^2} \left[ 1-\cos^{2p^2} \left(
\frac{\phi}{2} \right) \right],
\label{eq:p}
\end{equation}
Note that $V_{p=1}-J=-J\cos \phi=V(\phi)$. For $p>1$ the potential has a
smaller dip, and for $p<1$ a larger dip, as is shown in Fig.~\ref{fig:pot}.
Also note that for small $\phi$, the leading term $J\phi^2/2$ is quadratic
and is identical for $V$, $V_{\rm Villain}$, and $V_p$.

In the present paper, we will focus on the helicity modulus which measures
the resistance with respect to a uniform twist across the sample in one
direction. For a twist of size $\Delta_x$ across the $x$ direction, this
means that the Hamiltonian in the presence of the twist field is given by
$H(\Delta_x)= \sum_{\left< i, j \right>} V_p(\theta_i -
\theta_j-\Delta_x/L_x)$, where $L_x=L$ is the system size in the $x$ direction.
The helicity modulus $\Upsilon$ measures the increase of the free energy $F$
caused by the twist in the limit of small $\Delta_x$. This increment is
given by
$\frac{\partial^2 F}{\partial \Delta_x^2}|_{\Delta_x=0}
\Delta_x^2/2 \equiv \Upsilon \Delta_x^2/2$, where the
leading dependence in $\Delta_x$ is of the second order since a twist will
always increase the free energy or leave it invariant. For an interaction
potential $U(\phi)$, the helicity modulus is given by
\begin{equation}
\Upsilon= \left< e \right> - L^2 \beta \left< s^2 \right>,
\label{eq:hel}
\end{equation}
with $e \equiv L^{-2} \sum_{\left< ij \right>_x} U''(\theta_i - \theta_j)$ and
$s \equiv L^{-2} \sum_{\left< ij \right>_x} U'(\theta_i - \theta_j)$
where the sum is over all links in the $x$ direction. The derivatives are
with respect to the argument $\phi$ so that $U'(\phi) \equiv \partial U /
\partial \phi$ and $U'' \equiv \partial^2 U / \partial \phi^2$. We will here
also study a higher-order correlation function $\Xi \equiv \partial \Upsilon
/ \partial T = -\beta^2 \partial \Upsilon / \partial \beta$ written as
\begin{equation}
\Xi= \beta^2 \left\{ \left< e H \right> - \left< e \right> \left< H \right>
+ L^2 \left< s^2 \right> - L^2 \beta \left[ \left< s^2 H \right> - \left<
s^2 \right> \left< H \right> \right] \right\},
\label{eq:xi}
\end{equation}
and check its size dependence $\Xi \sim L^{\alpha_h}$ through
MC simulations.

\section{Monte Carlo Simulations}
\label{sec:mc}

\begin{figure}
\includegraphics[width=0.45\textwidth]{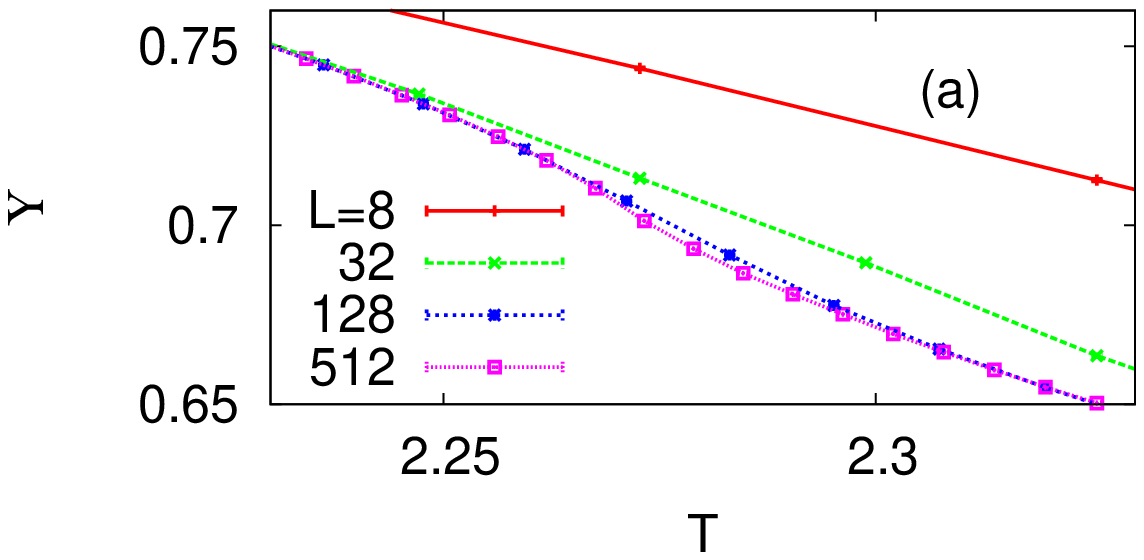}
\includegraphics[width=0.45\textwidth]{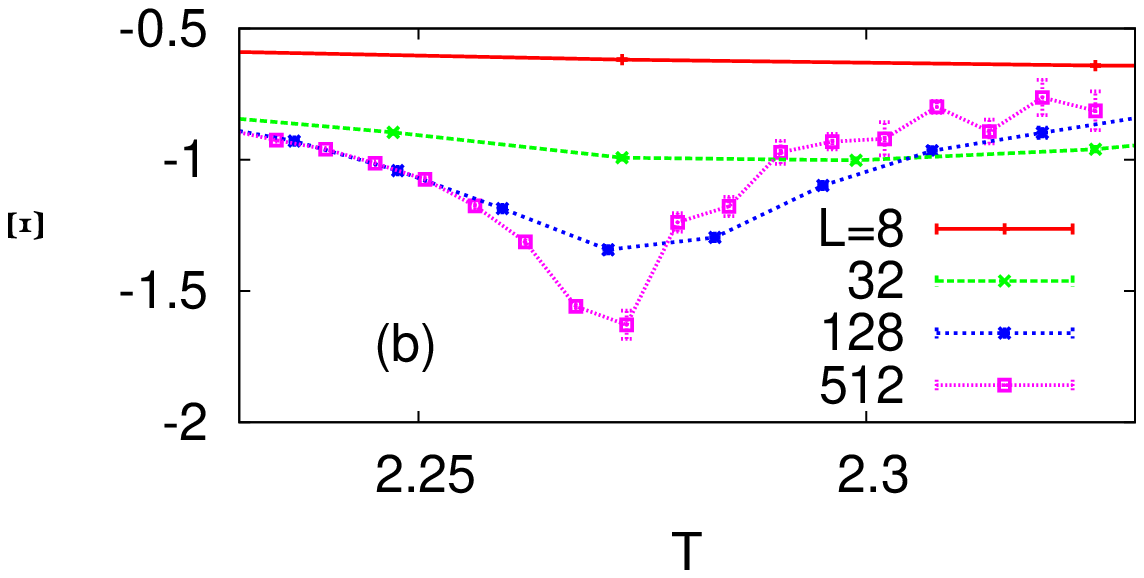}
\includegraphics[width=0.45\textwidth]{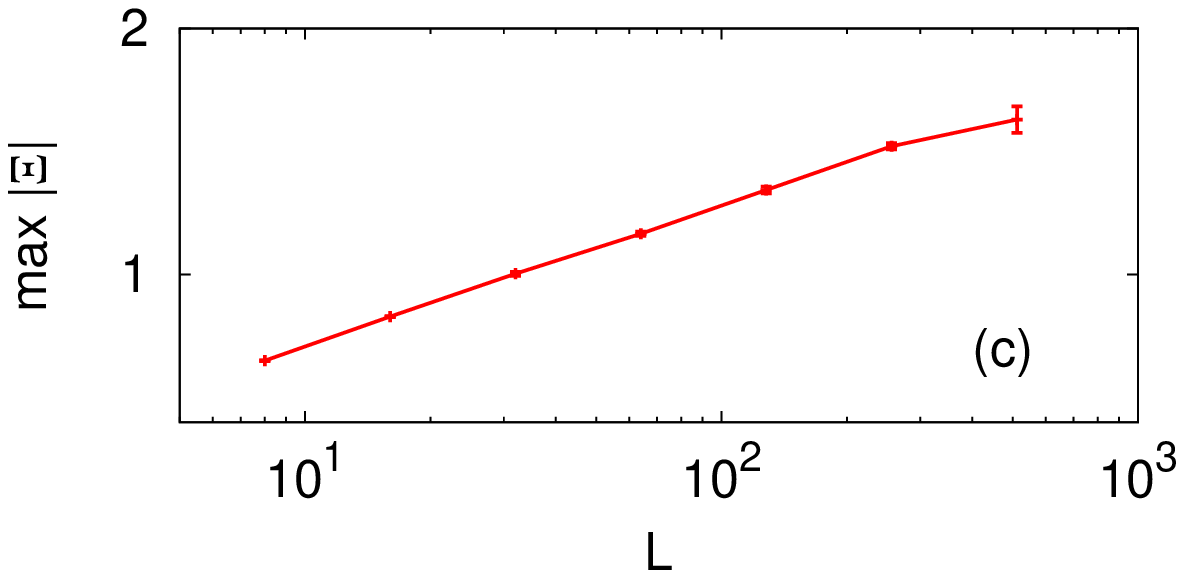}
\caption{(Color online) 2D Ising model ($q=2$). (a) Helicity modulus and (b)
its temperature derivative. (c) The peak heights of $\Xi$ shows a
logarithmic growth with the system size (see text).
Error bars are shown in all of these plots.}
\label{fig:ising}
\end{figure}

\begin{figure}
\includegraphics[width=0.45\textwidth]{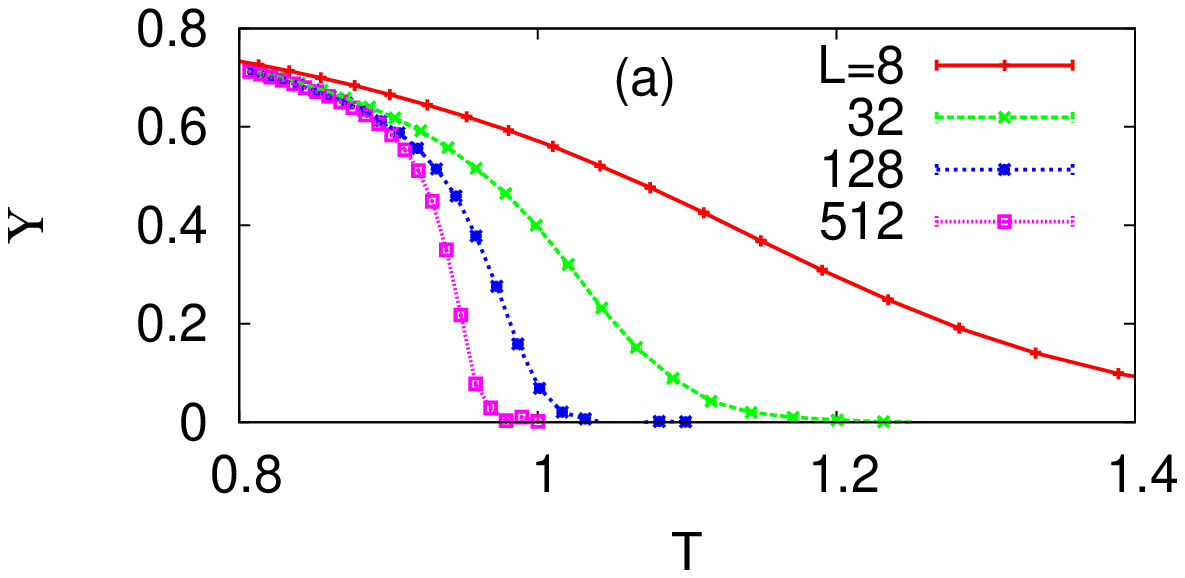}
\includegraphics[width=0.45\textwidth]{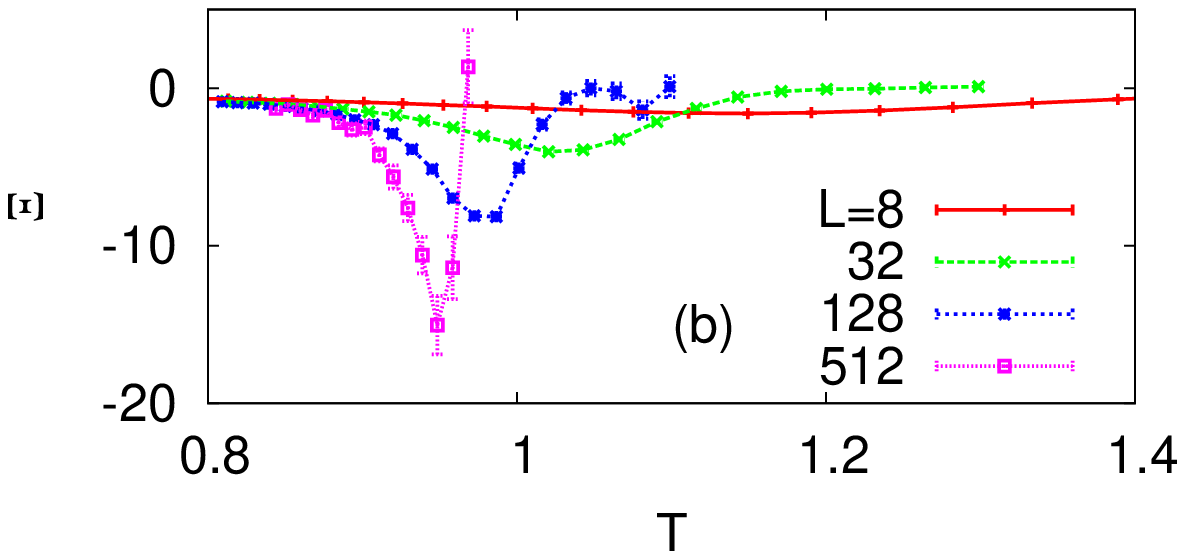}
\includegraphics[width=0.45\textwidth]{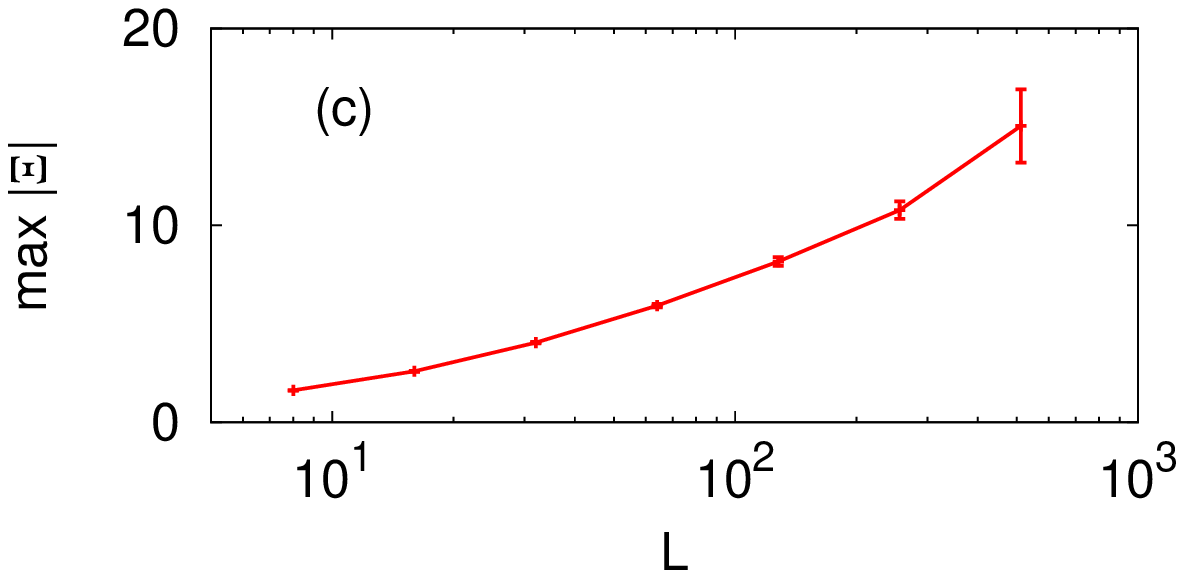}
\caption{(Color online) 2D $XY$ model ($q=\infty$). (a) Helicity modulus and
(b) its temperature derivative. (c) The peak heights of $\Xi$ shows a
divergence as $L$ grows. Error bars are also shown.}
\label{fig:xy}
\end{figure}

We use the Wolff single-cluster algorithm~\cite{wolff} throughout
this work. All the measurements presented here have been obtained
by updating clusters $O(10^6)$ up to $O(10^7)$ times after equilibration.
In Fig.~\ref{fig:ising}(a), we illustrate our calculations for
the two-state clock model, which corresponds to the 2D Ising model. At the
Ising transition, the helicity modulus makes a transition from a higher to a
lower value. However, the helicity modulus remains positive and nonzero for
all temperatures. This is in contrast to the KT transition for which the
helicity modulus is zero in the high-temperature phase. The Ising
criticality is instead reflected in the helicity modulus by the fact that
its temperature derivative diverges at the critical temperature. This
divergence is directly picked up by the correlation function $\Xi$, as
illustrated in Fig.~\ref{fig:ising}(b). From
the peak heights in Fig.~\ref{fig:ising}(b),
one can determine the size scaling $\Xi \sim
L^{\alpha_h}$ and the critical index $\alpha_h$. Figure~\ref{fig:ising}(c)
illustrates that $\alpha_h=0$ for the Ising model since peak heights are
consistent with a logarithmic growth. For the Ising model, this is a trivial
result since all correlations involving $s$ vanishes by $s=0$.
This mean that $\Xi$ is
proportional to the specific heat for the Ising model and hence
$\alpha_h=\alpha/\nu=0$.

\begin{figure}
\includegraphics[width=0.45\textwidth]{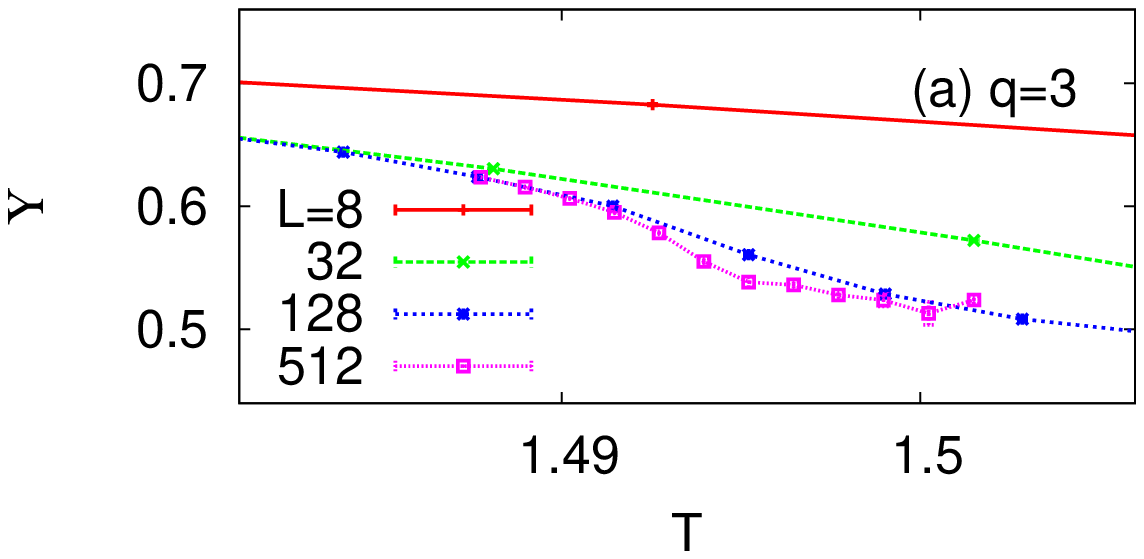}
\includegraphics[width=0.45\textwidth]{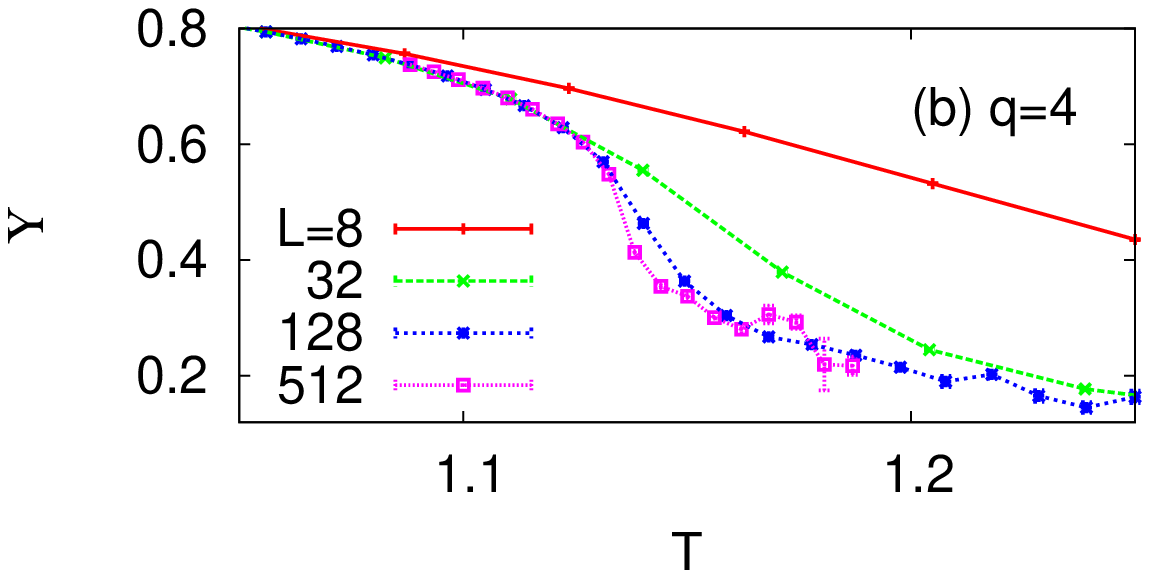}
\includegraphics[width=0.45\textwidth]{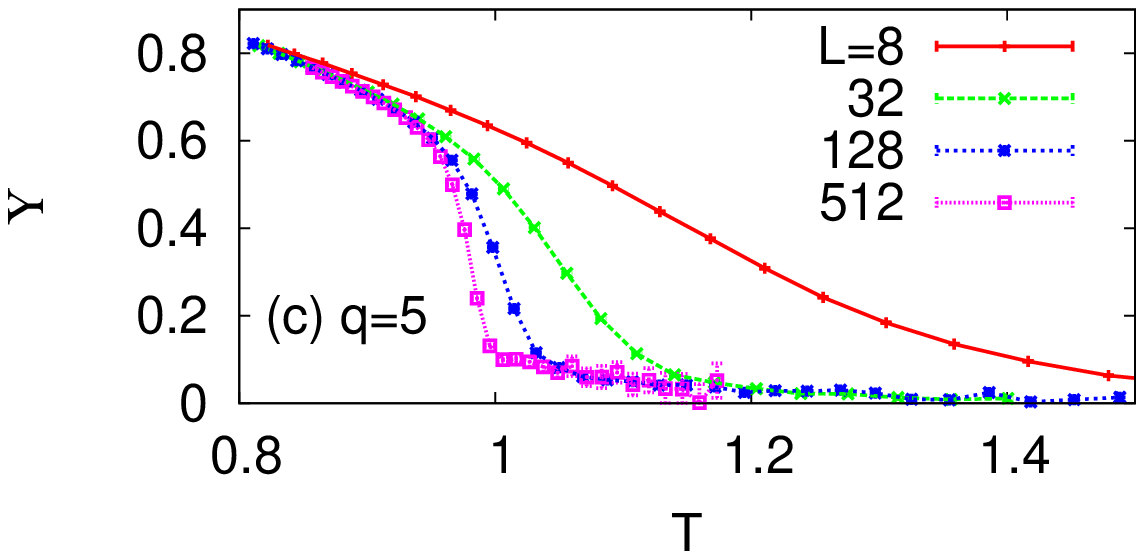}
\includegraphics[width=0.45\textwidth]{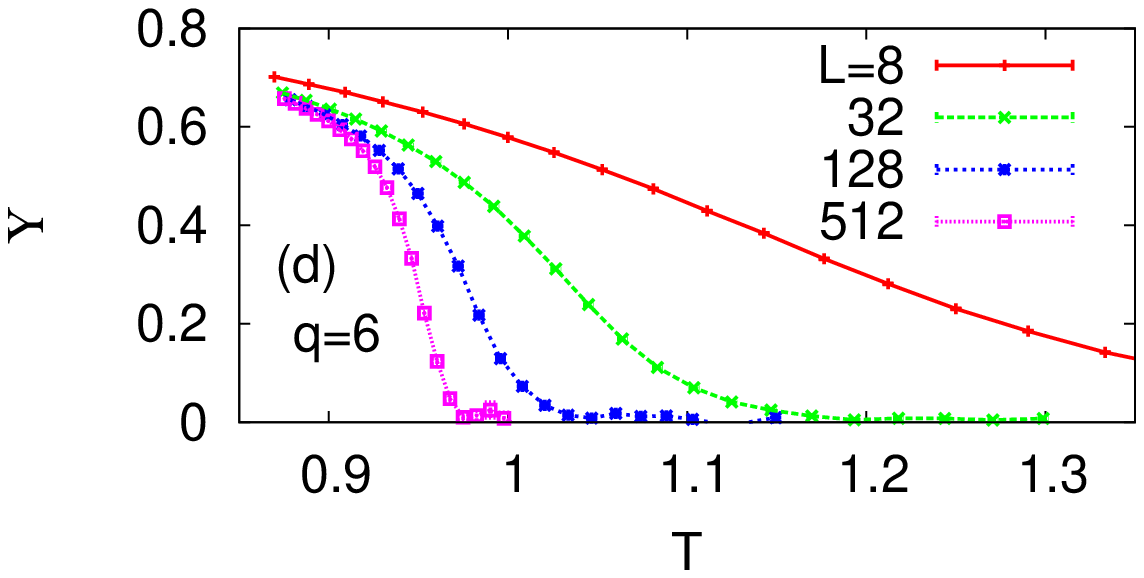}
\includegraphics[width=0.45\textwidth]{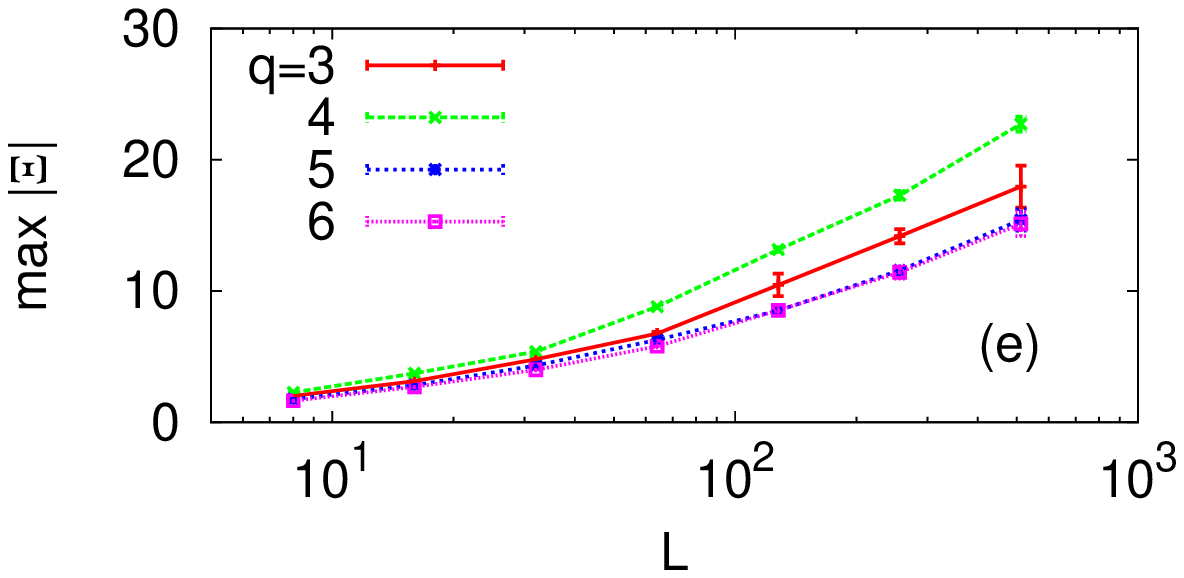}
\caption{(Color online) Helicity modulus measured for clock models with (a)
$q=3$, (b) $q=4$, (c) $q=5$, and (d) $q=6$, together with error bars.
(e) The peak heights of $\Xi$ at the $q$ values.}
\label{fig:clock}
\end{figure}

The other extreme for the clock models is the
limiting case $q=\infty$, which corresponds to the 2D $XY$ model, the
prototype model for the KT transition. Figure~\ref{fig:xy}(a) shows the
helicity modulus in the 2D $XY$ model for various sizes $L$. The
characteristics for the helicity modulus in case of the KT transition is
that it jumps from a finite value $2T_c/\pi$ to zero at $T_c$ in the limit
$L=\infty$~\cite{nelson,petter2}.
Figure~\ref{fig:xy}(a) is consistent with such a behavior.
Figure~\ref{fig:xy}(b) illustrates that the peak heights of $\Xi$ diverges
and the divergence is again shown to be consistent with a logarithmic
growth [Fig.~\ref{fig:xy}(c)]. The reason is, however,
different since the correlation length $\xi$ for the $KT$ transition
diverges as $\ln\xi \propto (T-T_c)^{-1/2}$~\cite{kos}.
At the same time, $\Xi$ is proportional to
$|T-T_c|^{-1/2}$~\cite{petter1}, leading to the size scaling $\Xi \sim
\ln L$. Our MC simulations suggest that $\Xi$ for all $q$ are consistent with
at least a logarithmic size divergence [see Fig.~\ref{fig:clock}(e)].
However, our MC simulations do not have enough precision at the largest
sizes to rule out a stronger divergence in a power-law form. The crucial point
in the present context is that the correlation function $\Xi$ displays a
phase-transition singularity, manifested in a size divergence, for
\emph{all} the clock models including the Ising and the $XY$ model.

\begin{figure}
\includegraphics[width=0.45\textwidth]{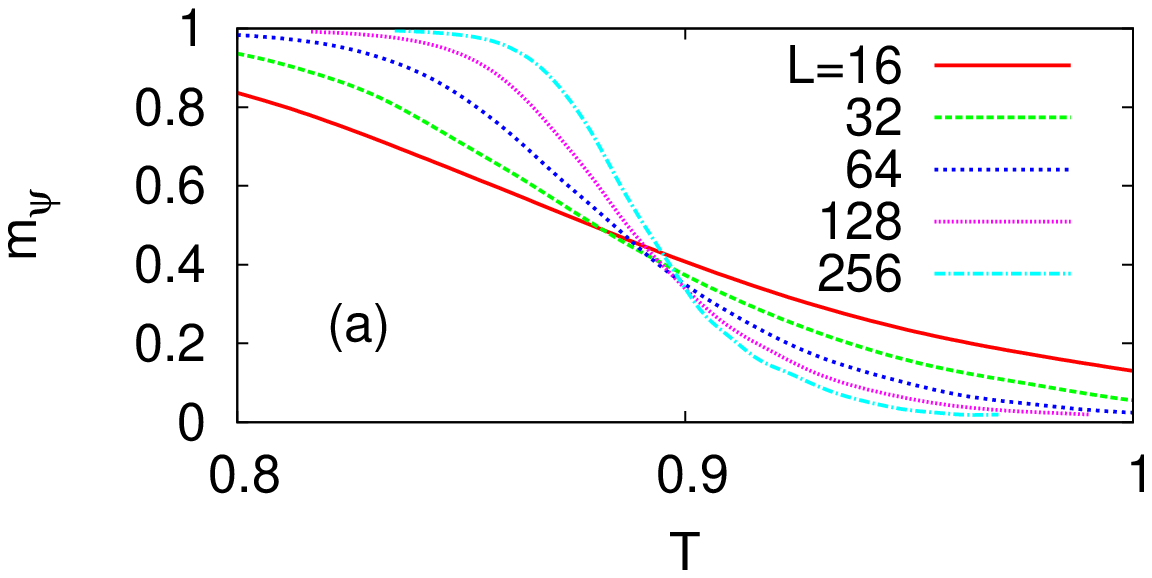}
\includegraphics[width=0.45\textwidth]{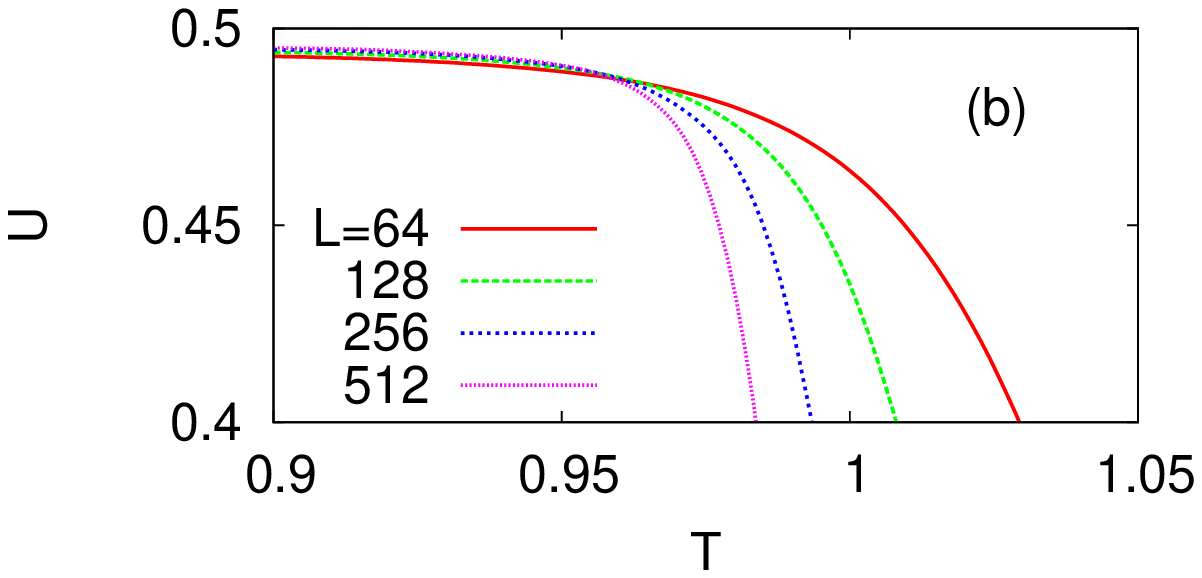}
\caption{(Color online) (a) The order parameter in Eq.~(\ref{eq:ang}) around
the lower transition temperature for the five-state clock model. Note that
the crossing points between curves move slightly to the right as the size
becomes larger. (b) Merging of Binder's cumulant around the higher
transition point for the same model.}
\label{fig:five}
\end{figure}

In Fig.~\ref{fig:clock}, we present the results for $q=3,4,5$, and $6$.
For each $q$, the helicity modulus $\Upsilon$ and the higher-order
correlation function $\Xi$ are given. Figure~\ref{fig:clock}(e) shows that
for all $q$, $\Xi$ has a critical divergence.
This means that in all cases there exists a phase transition which can be
associated with the helicity. An interesting point to note is that for
$q=2,3,4$, \emph{and} $5$, the helicity modulus itself remains finite for all
temperatures. This means that in these four cases the transition is
\emph{not} of the KT type. The clock models with $q=2,3$, and $4$ undergo
well-known phase transitions: for $q=2$ this is the Ising transition,
for $q=3$ the three-state Potts transition~\cite{wu} and for $q=4$ again an
Ising-like transition~\cite{suzuki}. These transitions go directly from
the low-temperature to the high-temperature phase, which rules out the
possibility for two consecutive transitions separated by a quasicritical phase
characterized by a power-law decay of spin correlations. However,
theoretical predictions strongly suggest that the five-state clock model does
have two transitions~\cite{jkkn,elitzur,domany1,cardy}.
This is also what we find from
our simulations: Fig.~\ref{fig:five}(a) determines the lower transition
using the order parameter introduced in Ref.~\cite{baek1},
\begin{equation}
m_{\psi} \equiv \left< \cos (q\psi) \right>,
\label{eq:ang}
\end{equation}
where $\psi$ means the phase of the magnetization vector so that $\mathbf{m}
\equiv L^{-2} \sum_j e^{i \theta_j} = |m| e^{i\psi}$. This parameter
distinguishes the true long-range order from the quasi-long-range order. On
the other hand, Fig.~\ref{fig:five}(b) detects the upper transition, at
which the spin-correlations start to decay exponentially, using Binder's
cumulant,
\begin{equation}
U \equiv 1 - \frac{\left< |m|^4 \right>}{2 \left< |m|^2 \right>^2}.
\label{eq:binder}
\end{equation}
These measurements show that there are indeed two separated transitions.
Thus the conclusion for the five-state clock model is that it does
have two consecutive transitions. However, contrary to the theoretical
expectations in Refs.~\cite{jkkn,elitzur,domany1,cardy}, it is not the KT
transition since as is illustrated in Fig.~\ref{fig:clock}(c), the helicity
modulus remains finite for all temperatures precisely as for $q=2,3$, and
$4$. Nor is it a discontinuous transition since we do not observe any double
peaks in the energy distribution~\cite{baek1}.

\begin{figure}
\includegraphics[width=0.45\textwidth]{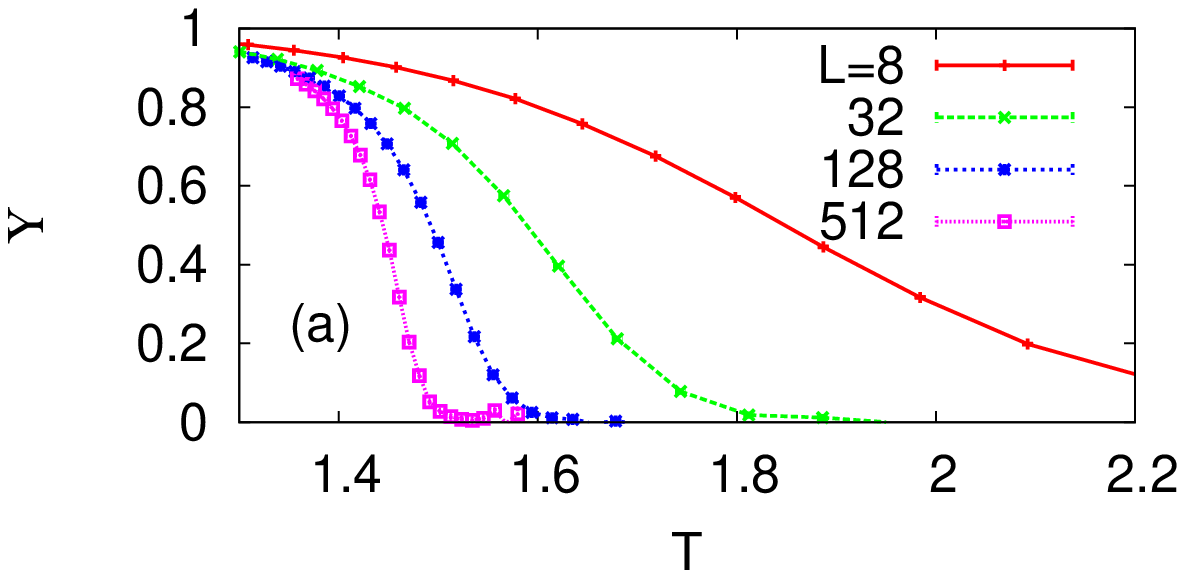}
\includegraphics[width=0.45\textwidth]{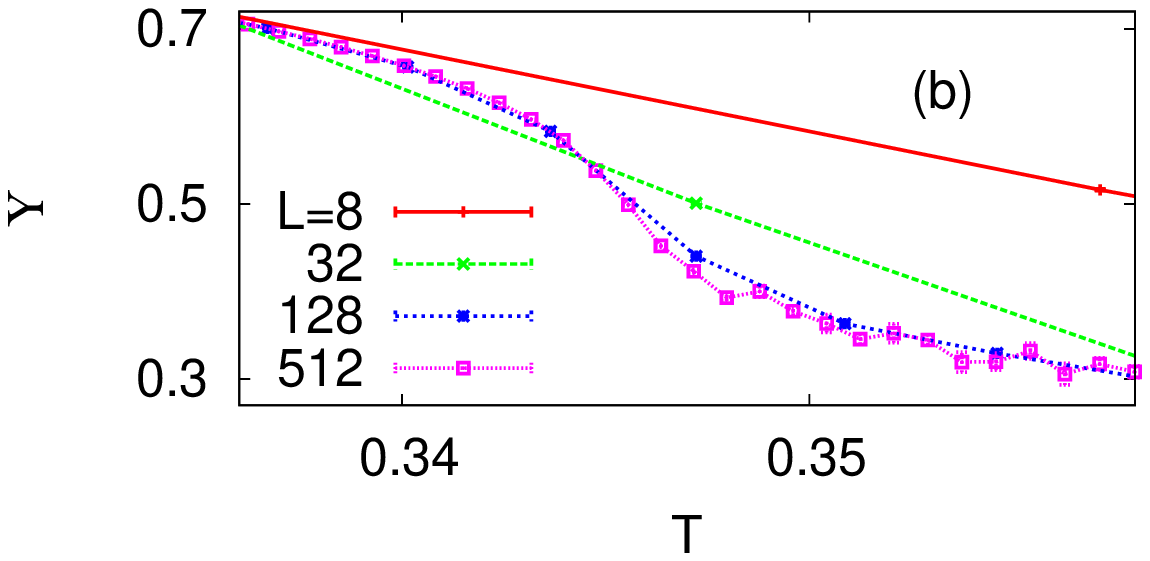}
\includegraphics[width=0.45\textwidth]{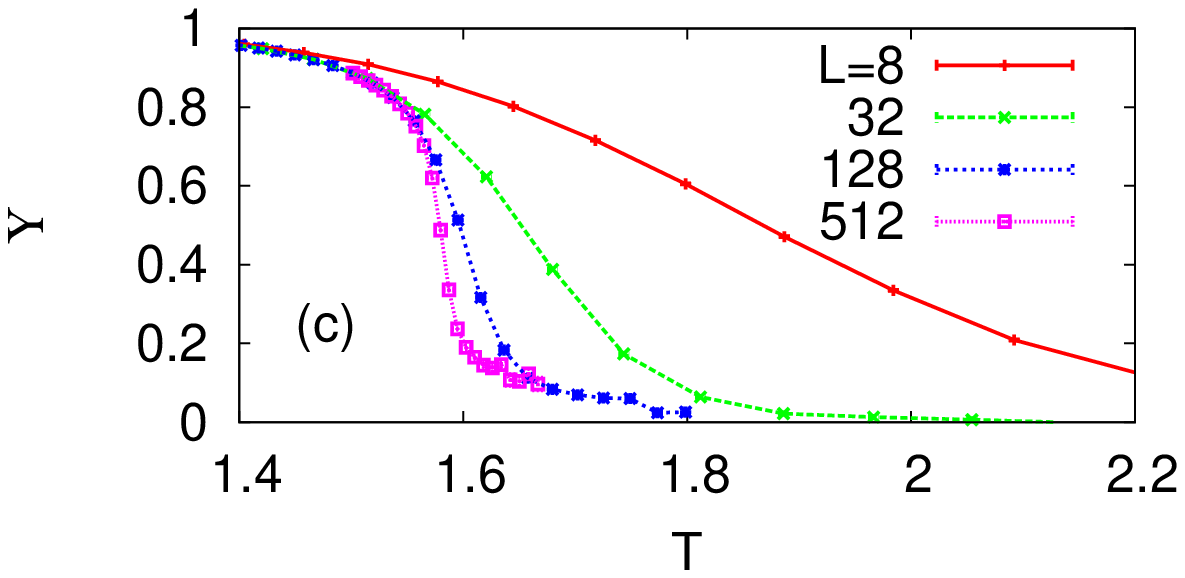}
\caption{(Color online) The helicity modulus for (a) the five-state Villain
model, (b) the generalized six-state clock model with $p=2.0$, and (c) the
four-state Villain model. Error bars are shown but usually smaller than
the symbol sizes.}
\label{fig:di}
\end{figure}

In order to better understand the reason for this, we first note that the
conclusions in Refs.~\cite{jkkn,elitzur,cardy} were obtained using the Villain
approximation. In Fig.~\ref{fig:di}(a), we show the result for the helicity
modulus for the five-state clock model using the Villain potential (compare
Fig.~\ref{fig:pot}). In this case,
the helicity modulus does indeed vanish. This means that, in accordance with
Refs.~\cite{jkkn,elitzur,domany1,cardy}, the five-state clock model within
the Villain
approximation does have two consecutive transitions where the higher
one is the KT transition. The crucial point made here is that this is not
true for the real five-state clock model. The reason for this change in the
transition is clearly the interplay between the number of clock states, $q$,
and the detailed shape of the potential. This is further illustrated in
Fig.~\ref{fig:di}(b), which shows the helicity modulus for the six-state
clock model using the generalized potential in Eq.~(\ref{eq:p}) with $p=2.0$,
which gives a slightly flatter potential (Fig.~\ref{fig:pot}).
For this potential, the six-state clock model has two consecutive
transitions where the helicity modulus does not
vanish at the upper transition, just as for the five-state clock model with the
usual cosine potential. However, the six-state clock model has two consecutive
transitions where the upper one is the KT transition both for the usual cosine
potential [Fig.~\ref{fig:clock}(d)] and for the Villain potential.
Figure~\ref{fig:di}(c) shows the helicity modulus for the four-state clock
model within the Villain approximation. However, in this case, the helicity
modulus remains nonzero for all temperatures as for the usual cosine
potential.

\section{Conclusions}
\label{sec:con}

From numerical simulations, it was verified that the five-state clock
model has two transitions. It was also found that the helicity modulus
$\Upsilon$ does not vanish at the upper transition. Yet it was found that
the helicity modulus does vanish at the upper transition provided the
cosine potential is replaced by the Villain potential. In this latter case,
the transition is perfectly consistent with the standard KT transition.
Heuristically, this transition can be interpreted in terms of the
unbinding of vortex-antivortex pairs: below the transition temperature,
these pairs are bound together whereas above it the effective
vortex-anivortex attractive interaction is softened such that free unbound
vortices appear. It is the presence of such free vortices which is signaled
by the vanishing of $\Upsilon$~\cite{nelson,petter2,petter1}.
It seems plausible that a description of the
phase transition in terms of vortex-antivortex pairs will remain adequate
even after the tiny change from the Villain potential to the usual
clock-model potential. If so, this means that the rapid drop of $\Upsilon$ at
the transition for the five-state clock model is again associated with a
softening of the effective vortex-antivortex interaction, with the
important difference that this softening is not enough to create free
vortices. In this interpretation, it is the lack of free vortices which
forces $\Upsilon$ to remain finite for all temperatures. Whether or not
the softening will be enough to create free vortices then depends on an
intricate balance between the number of clock states, $q$, and the precise
form of the potential. This interpretation is consistent with the finding
that, whereas the six-state clock model undergoes the KT transition for
which the helicity modulus vanishes, it only takes a small modification of
the potential using the systematic parameterization in Eq.~(\ref{eq:p}) to
change the transition such that the helicity modulus does not vanish. From
this perspective, the topological transition of the five-state clock
model, for which the helicity modulus does not vanish, appears to be a
new and weaker cousin of the proper KT transition.

\acknowledgments
We acknowledge the support from the Swedish Research Council
with the Grant No. 621-2002-4135.
This research was conducted using the resources of High Performance
Computing Center North (HPC2N).

%\bibliographystyle{revtex}
%\bibliography{clock}

\end{document}